\documentclass[apjl]{emulateapj}

\usepackage{hyperref}
\usepackage{amsmath}
\usepackage{amssymb}
\usepackage{graphicx}
\usepackage{color}
\usepackage{natbib}

\def\be{\begin{equation}}
\def\ee{\end{equation}}
\def\ba{\begin{eqnarray}}
\def\ea{\end{eqnarray}}

\begin{document}

\title{An alternative interpretation of GW190412 as a binary black hole merger with a rapidly spinning secondary}
\shorttitle{GW190412 spins}

\author{Ilya Mandel\altaffilmark{1,2,3}}
\affil{$^1$Monash Centre for Astrophysics, School of Physics and Astronomy, Monash University, Clayton, Victoria 3800, Australia}
\affil{$^2$The ARC Center of Excellence for Gravitational Wave Discovery -- OzGrav}
\affil{$^3$Institute of Gravitational Wave Astronomy and School of Physics and Astronomy, University of Birmingham, Birmingham, B15 2TT, United Kingdom}
\email{ilya.mandel@monash.edu}
\author{Tassos Fragos\altaffilmark{4}}
\affil{$^4$Geneva Observatory, University of Geneva, Chemin des Maillettes 51, 1290 Sauverny, Switzerland}

\begin{abstract}
The LIGO-Virgo collaboration recently reported the properties of GW190412, a binary black hole merger with unequal component masses (mass ratio $0.25^{+0.06}_{-0.04}$ when using the EOBNR PHM approximant) and a non-vanishing effective spin aligned with the orbital angular momentum.  They used uninformative priors to infer that the more massive black hole had a dimensionless spin magnitude between 0.17 and 0.59 at 90\% confidence.  We argue that, within the context of isolated binary evolution, it is more natural to assume a priori that the first-born, more massive black hole has a negligible spin, while the spin of the less massive black hole is preferentially aligned with the orbital angular momentum if it is spun up by tides.  Under this astrophysically motivated prior, we conclude that the lower mass black hole had a dimensionless spin component between 0.64 and 0.99 along the orbital angular momentum.
\end{abstract}

\maketitle

\section{Introduction}

Gravitational waves from the coalescence of two black holes, GW190412, were detected by the advanced LIGO \citep{AdvLIGO} and Virgo \citep{AdvVirgo} gravitational-wave observatories on 12 April, 2019 \citep{GW190412}.  The data were analysed with a range of gravitational waveform models.  In all cases, the data point to a significantly asymmetric mass ratio, with a heavier (primary) black hole of roughly 30 solar masses and a lighter (secondary) black hole of less than 10 solar masses.  Moreover, this is one of the few systems detected so far with a clearly non-negligible spin.  The effective spin is defined as
\begin{equation}\label{eq:chieff}
\chi_\mathrm{eff} \equiv \frac{(m_1 \vec{\chi_1} + m_2 \vec{\chi_2}) \cdot \hat{L}}{m_1+m_2},
\end{equation}
where $m_1$ and $m_2$ are the primary and secondary masses, $\chi_1$ and $\chi_2$ are the corresponding dimensionless spins ($0 \leq \chi_i \equiv |\vec{\chi_i}| \leq 1$) and $\hat{L}$ is the unit vector along the orbital angular momentum.  It is inferred to be $\chi_\mathrm{eff} = 0.28^{+0.07}_{-0.08}$ (here and below, median and bounds of 90\% credible interval) when using the EOBNR PHM (effective one-body calibrated to numerical relativity incorporating precession and higher multipoles) approximant \citep{Ossokine:2020}. We quote results for this approximant throughout this paper.

Unfortunately, this is the only spin combination that is readily measurable with gravitational-wave observations.  A measure of the spin component in the orbital plane, $\chi_\mathrm{p}$, is very poorly constrained: \citet{GW190412} report that there is only tentative evidence for a non-zero $\chi_\mathrm{p}$, as the signal-to-noise ratio of the excess power due to precession overlaps with the expectation from random noise fluctuations (see figure 6 of \citealt{GW190412}).  While high values of $\chi_\mathrm{p}$ are disfavored by the data, low values only appear to be ruled out by the chosen prior (see figure 5 of \citealt{GW190412}).  It is impossible to directly measure the magnitudes of individual spins or their tilt angles relative to the orbital angular momentum from the combination $\chi_\mathrm{eff}$ alone.  Any such attempts are therefore very sensitive to a priori assumptions about the plausible distributions of these properties, which enter the analysis as priors.

\citet{GW190412} assumed priors that were uniform in component spin magnitudes $\chi_i$ and isotropic in the directions, i.e., uniform in the cosine of the tilt angles $\hat{\chi_i} \cdot \hat{L}$.  From this, they infer that the primary must be relatively rapidly spinning, with a spin magnitude $\chi_1 = 0.46^{+0.12}_{-0.15}$, while the spin of the secondary is not constrained.  We argue in section II that our best understanding of stellar and binary evolution suggests a different prior, in which the primary black hole is not significantly spinning while the secondary's spin may be large but must be  preferentially aligned with the orbital angular momentum.  In section III, we re-weight the posterior samples of \citet{GW190412} to infer that the secondary's spin has a projection of $\vec{\chi_2} \cdot \hat{L} = 0.88^{+0.11}_{-0.24}$ along the orbital angular momentum.  We conclude with a brief summary of future prospects in section IV.


\section{Astrophysical prior}

In the last several years, various teams have argued that the first-born black hole in an isolated merging binary black hole system is likely to have very slow spin as long as there is efficient angular momentum transport within the star, while the secondary may be rapidly spinning only if the binary is sufficiently tight before the secondary's collapse to enable tidal locking \citep[e.g.,][]{Kushnir:2016,HotokezakaPiran:2017,Zaldarriaga:2017,Qin:2018,Belczynski:2020,FullerMa:2019,Bavera:2019}.  Given the uncertainties in the detailed models, here we attempt to pedagogically lay out the key arguments and caveats.

We restrict our discussion to isolated binaries.  Other binary black hole formation channels are plausible, including formation in dense stellar environments and in hierarchical three-body systems \citep[see, e.g.,][for a review]{MandelFarmer:2018}.  However, since the classical isolated binary evolution channel is consistent with the rate and property distribution of all events observed so far \citep[e.g.,][]{Neijssel:2019}, we assume that GW190412 was also formed through this channel.  We further assume that the more massive black hole is formed first from the initially more massive and more rapidly evolving star; while the mass ratio can be reversed during binary evolution, the very asymmetric mass ratio of GW190412 is unlikely to arise from a binary that experienced mass ratio reversal (but may point to formation in a lower-metallicity environment, \citealt{Stevenson:2017}).

It is relatively easy to spin up a star or stellar core.  Assuming a star of mass $M$ and radius $R$ accretes a mass $m$ of material that is moving at a Keplerian velocity $v=\sqrt{G M / R}$ at the star's equator, it gains an angular momentum $S = m R v = m \sqrt{G M R}$.  If the star is initially non-spinning, it will end up with a dimensionless spin magnitude 
\begin{equation}\label{eq:spinup}
\chi \equiv \frac{c S}{G M^2} = \frac{m}{M} \sqrt{\frac{R c^2}{G M}}.
\end{equation}  
The second term has a value of order $10^{3}$ for a typical star; therefore, only a fraction of a percent of the star's mass needs to be accreted in order to significantly spin up the star.

On the other hand, once a black hole is formed, spinning it up requires a significant amount of accreted material.  Ignoring the complexities of general relativity and naively applying equation (\ref{eq:spinup}) provides a simple back-of-the-envelope estimate.  The second term in equation (\ref{eq:spinup}) has a value of order unity for a black hole.  This indicates that the black hole mass must be roughly doubled to produce a rapidly spinning object.  A more accurate calculation \citep[e.g.,][]{Thorne:1974ve,KingKolb:1999,Podsiadlowski:2002,Fragos:2015} shows that 20\% of the black hole's mass must be added to bring the spin of a non-spinning black hole to 0.5, and the mass must be more than doubled to bring the spin to 0.99.   The mass doubling timescale for a black hole accreting at the Eddington limit is of order 100 million years -- far longer than the $\lesssim 10^5$ year lifetime of high-mass X-ray binaries composed of a black hole and its wind-shedding companion.  Although \citet{MorenoMendez:2008} argued that hypercritical accretion may allow a significant amount of matter to be accreted at super-Eddington rates, numerical simulations indicate that only a small fraction of the black hole's mass is likely to be accreted during the common envelope phase \citep{MacLeodRamirezRuiz:2015,De:2019}. Therefore, any angular momentum in a black hole must come from the progenitor star or the supernova itself.

\begin{figure}
\centering
\includegraphics[width=0.95\columnwidth]{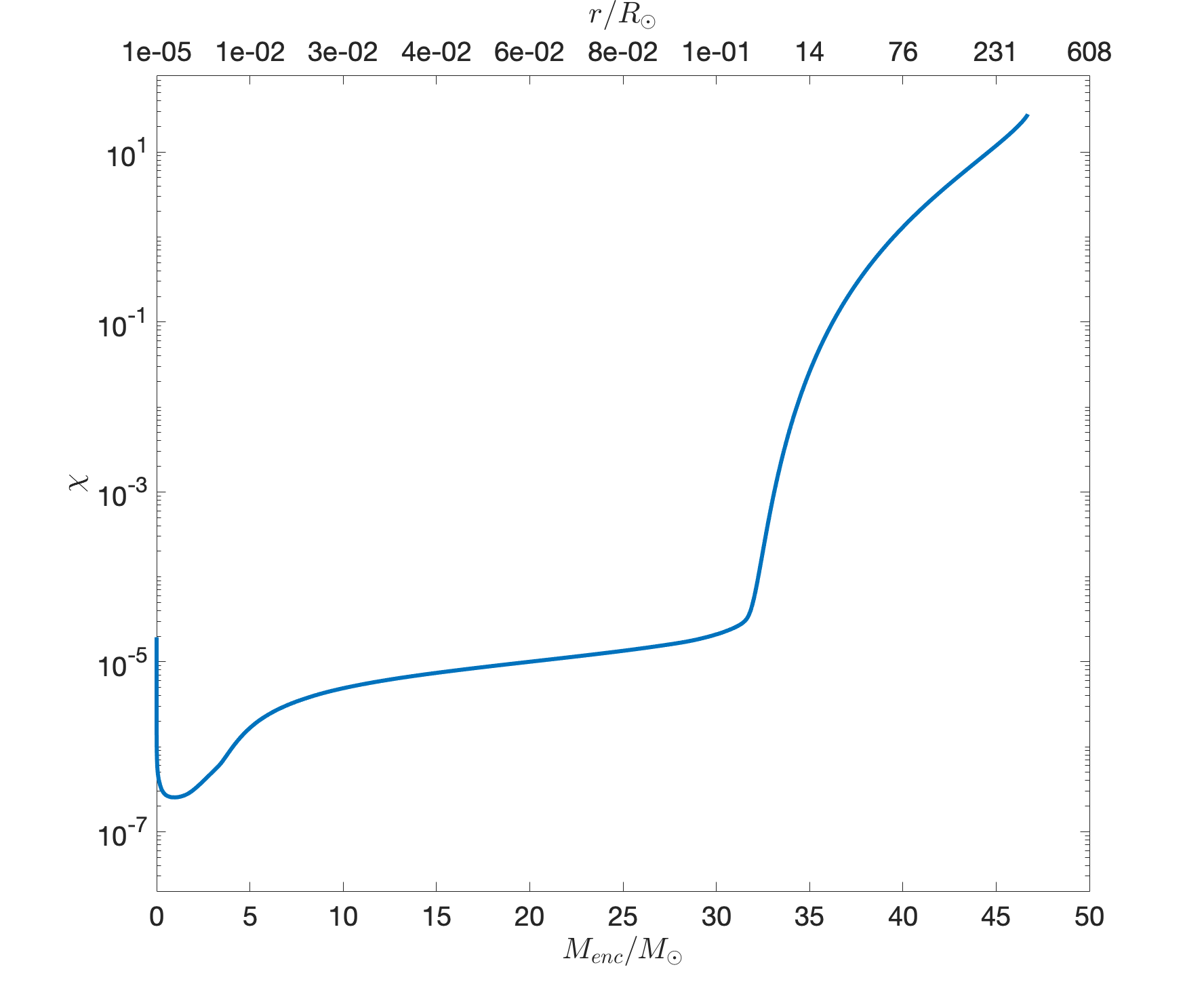}
\caption{Enclosed dimensionless spin as a function of the mass coordinate for a maximally rigidly  rotating giant with an initial mass of $60 M_\odot$ at metallicity $Z=0.0002$ evolved until just before collapse with MESA \citep{Paxton:2011}.  The top abscissa shows the radius of the given mass coordinate.  Although the total $\chi \gg 1$, almost all of the angular momentum is in the envelope, and the core has a dimensionless angular momentum $\chi \ll 1$.}\label{figure:chiMenc}
\end{figure}

In order for gravitational-wave emission to bring a binary to coalesce within the age of the Universe, the black holes must be quite close -- about 30 solar radii apart for the masses of GW190412 \citep{Peters:1964}.  This is much smaller than the size of the evolved supergiant progenitors of these black holes, so the stars must interact with several episodes of mass transfer, possibly including a common-envelope phase.  Consequently, the envelopes of the stars will be stripped off.  But the envelopes contain the bulk of the moment of inertia, and hence, with efficient angular momentum transport, the bulk of the angular momentum.  For example, figure \ref{figure:chiMenc} shows the dimensionless spin enclosed within a given mass coordinate for a maximally rotating giant star under the assumption of rigid body rotation; while the total dimensionless spin exceeds 1, the spin of the compact core is 5 orders of magnitude lower -- i.e., differential rotation by more than 6 orders of magnitude would be required for the stripped core to retain a high spin parameter.  This extreme differential rotation appears inconsistent with asteroseismic observations of (albeit, much less massive) giants \citep{Cantiello:2014,denHartogh:2019} and with theoretical models of angular momentum transport \citep[][but see \citealt{Eggenberger:2019}]{Tayler:1973,Spruit:2002,Fuller:2019}.  We therefore conclude that even when stars are rapidly rotating earlier in their evolution, the stellar core left behind after stripping will have negligible spin.

Binary black hole formation through chemically homogeneous evolution \citep{Marchant:2016,MandeldeMink:2016,deMinkMandel:2016,duBoisson:2020} avoids angular momentum loss through envelope stripping, and can therefore yield two spinning black holes.  However, it is expected to give rise to black-hole binaries with large, nearly equal masses; therefore, GW190412 is unlikely to have evolved through this channel.  

It is also possible that black holes could spin up during the supernova itself.  This could happen through the loss of angular momentum during the explosion, with the remnant acquiring the conjugate angular momentum.  However, black holes as massive as 30 $M_\odot$ are likely to form through complete fallback \citep{Fryer:2012,Adams:2017}.  Alternatively, \citet{Batta:2017,Schroeder:2018} proposed that black holes could spin up during the supernova if some of the ejecta were torqued by the companion before falling back, transferring part of the orbital angular momentum to the black hole; however, this generally requires some of the ejecta velocities to be comparable to the orbital velocities, and is unlikely to lead to significant spin-up in most cases.   Instabilities such as the standing accretion shock instability during core collapse \citep{BlondinMezzacappa:2007} or gravity waves during O and Si shell burning \citep{Fuller:2015} have also been proposed as mechanisms for stochastically transferring angular momentum to the core.  While this angular momentum transfer could be significant for neutron stars, it is not expected to be sufficient to create any measurable spin for black holes \citep[e.g.,][]{MorenoMendez:2016}.

This leaves tides as the most likely mechanism to spin up the star.  These tides must operate after stripping -- otherwise, even if the star is tidally spun up earlier in its evolution (say, on the main sequence), it will still lose the vast majority of its angular momentum when its envelope is removed during subsequent mass transfer.  However, tides are only efficient in close binaries, because the tidal efficiency scales as a high power of the ratio of the star's radius to the binary separation \citep[e.g.,][]{Hut:1981}.   Stripped stars are compact, with radii of order one solar radius, so tides can only spin up stripped stars if the binary is very compact indeed, with separations of only a few solar radii.  A binary cannot be this compact when the progenitor of the primary is a stripped star and the progenitor of the secondary is on the main sequence, since a main-sequence massive companion would not fit into a binary this tight. Hence, we do not expect the primary to be rapidly spun up by tides.  

One noteworthy exception could be a double-core common envelope event, which simultaneously strips both the primary and the secondary of their envelopes and brings the remnant cores close together \citep{BetheBrown:1998,Dewi:2006}.  In this case, which requires further investigation, it may be possible for both companions to be tidally spun up.  However, such events require similar companion masses so that both stars are evolved off the main sequence at the time of the interaction, which again makes it implausible as the formation channel of GW190412.

It may be possible that mass transfer during the main sequence will simultaneously spin up the donor through tides and remove enough of its envelope to prevent subsequent expansion of the star and loss of angular momentum through envelope stripping in the giant phase.  This model was proposed by \citet{Qin:2019} to explain the very high observed black hole spins in some high-mass X-ray binaries, particularly Cygnus X-1, LMC X-1 and M33 X-7 (but see \citealt{MillerMiller:2015,Kawano:2017} for a discussion of the uncertainties in spin measurements).  However, these high-mass X-ray binaries likely form a distinct population from merging compact-object binaries \citep{HotokezakaPiran:2017} and will not merge within the age of the Universe \citep{Belczynski:2012HMXB}, being too wide to merge at their present separations and too tight to survive another common envelope.

Thus, tidal spin-up is likely to operate only on the secondary, if mass transfer -- most likely during a common-envelope event -- brings its stripped core sufficiently close to the already formed black hole.  This is the primary channel for tidally spinning up the secondary considered by \citet{Kushnir:2016,Zaldarriaga:2017,Belczynski:2020,Qin:2018,Bavera:2019}.  Some of these papers treated tides in a simplified way, assuming the stripped He star to be fully synchronized with the orbit if the estimated synchronization timescale is shorter than the wind mass-loss timescale and non-spinning otherwise. Others used detailed binary evolution models that include, e.g., orbital widening and spin down through wind-driven mass loss and angular momentum transport in the stellar interior.  Although quantitatively the results of these studies differ, they all conclude that the primary is likely to have negligible spin in the presence of efficient angular momentum transport, while the secondary may be at least partially tidally spun up in a subset of merging binaries.   

We are thus left with a generic model for the formation of a merging binary black hole from an isolated binary in which the primary has negligible spin while the secondary may be partially spun up.  The secondary may have experienced a supernova natal kick, and there is some evidence from both models and observations that relatively low-mass black holes may experience kicks of a few tens to $~\sim100$ km s$^{-1}$ \citep{Willems:2005,Fragos:2009,Fryer:2012,Mueller:2016,Mandel:2015kicks,Mirabel:2016,WyrzykowskiMandel:2019,Atri:2019} and possibly even a few hundred km s$^{-1}$ \citep{Repetto:2017}.  However, as argued above, the secondary will only be tidally spun up in very close binaries, where its orbital velocity is likely to approach a thousand km s$^{-1}$, so such kicks will not lead to significant spin-orbit misalignment.  In any case, we can consider only inference on the dimensionless spin component of the secondary along the direction of the orbital angular momentum, $\vec{\chi_2} \cdot \hat{L}$.  Therefore, our chosen prior is: {\it negligible spin for the primary; and a uniform on $[0,1]$ prior for $\vec{\chi_2} \cdot \hat{L}$}, to reflect our ignorance.  The latter is partly inspired by figure 8 of \citet{Bavera:2019}, but as we will see, the data support a high value of $\vec{\chi_2} \cdot \hat{L}$ and the exact shape of the prior on this quantity is not as important as the constraint on the primary spin.  While we could also use binary population synthesis models to place priors on the masses, we choose to follow the broad, uniform in component mass priors of \citet{GW190412} for simplicity, as the masses are relatively well measured and thus insensitive to other broad prior choices. 

\section{Results}

\begin{figure}
\centering
\includegraphics[width=0.95\columnwidth]{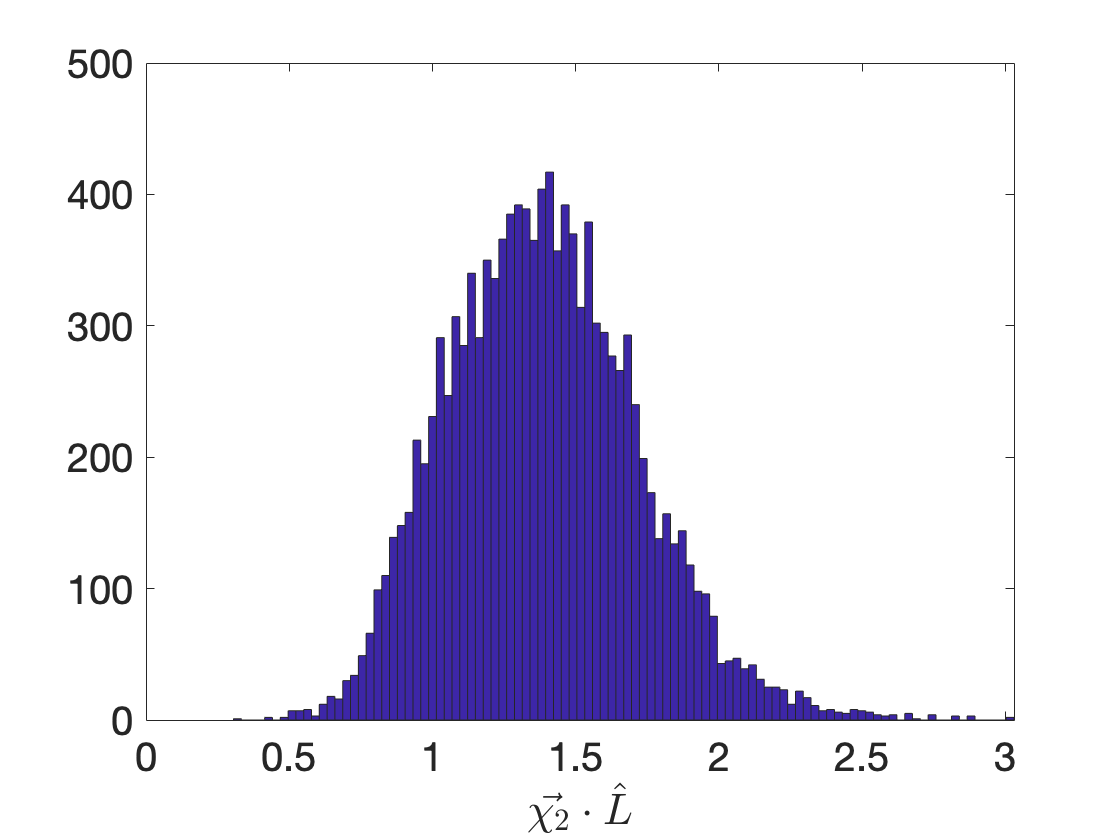}
\caption{$\chi_\mathrm{eff}$ posterior samples from \citet{GW190412} transformed into $\vec{\chi_2} \cdot \hat{L}$ according to equation (\ref{eq:chi2}) without prior re-weighting.}\label{figure:spin2z}
\end{figure}

Before proceeding with imposing our alternative prior, we consider a simple re-interpretation of the existing posterior on $\chi_\mathrm{eff}$ under the assumption that the effective spin is due entirely to the secondary.  If $\chi_1=0$, equation (\ref{eq:chieff}) can be re-written as $\chi_\mathrm{eff} = m_2 \vec{\chi_2} \cdot \hat{L} / (m_1 + m_2)$, or  
\begin{equation}\label{eq:chi2}
\vec{\chi_2} \cdot \hat{L} = \frac{m_1+m_2}{m_2} \chi_\mathrm{eff}.
\end{equation}  
Applying this transformation directly to the posterior samples supplied at \url{https://dcc.ligo.org/LIGO-P190412/public}, we obtain the histogram shown in figure \ref{figure:spin2z}.  While this transformation extends $\vec{\chi_2} \cdot \hat{L}$ into the unphysical regime above 1, figure \ref{figure:spin2z} shows that, even without adjusting the priors, part of the inferred $\chi_\mathrm{eff}$ space is consistent with a non-spinning primary and a rapidly spinning secondary.

Our goal is to recompute the posterior using the prior specified at the end of section II.  Given the computational cost of gravitational-wave inference \citep{Veitch:2014}, this is most efficiently done by re-weighting the existing posterior samples with a new prior.  If the prior used in \citet{GW190412} on source parameters $\vec{\theta}$ is $\pi_\mathrm{LVC}(\vec{\theta})$ and the likelihood of observing the data $d$ given $\vec{\theta}$ is $p(d|\vec{\theta})$, then the \citet{GW190412} posterior on the parameters $\vec{\theta}$ given data $d$ is
$$p_\mathrm{LVC} (\vec{\theta}|d) = \frac{ \pi_\mathrm{LVC}(\vec{\theta}) p(d|\vec{\theta})}{p(d)}.$$
We can then obtain the posterior using an alternative prior $\pi(\vec{\theta})$ as
\begin{equation}\label{eq:reweight}
p(\vec{\theta}|d) = \frac{ \pi(\vec{\theta}) p(d|\vec{\theta})}{p(d)}
= w(\vec{\theta}) p_\mathrm{LVC} (\vec{\theta}|d),
\end{equation}
where the weights are just the ratio of the two priors, $w(\vec{\theta}) = \pi(\vec{\theta}) / \pi_\mathrm{LVC}(\vec{\theta})$.

In principle, we should re-weigh all samples according to our desired prior.  In practice, this is impossible, because there are no samples in the set of measure zero with $\chi_1=0$.  Even if we replace that delta function prior with a somewhat broader one -- say, a Gaussian centered on zero -- we still suffer from too few samples: only one of the 11992 posterior samples provided for the EOBNR PHM approximant has $\chi_1 < 0.1$.  This is not surprising: there is a much greater parameter space within the $\pi_\mathrm{LVC}$ prior to obtain the desired value of $\chi_\mathrm{eff}$ through larger values of $\chi_1$, which can be consistent with almost any $\chi_2$, than through a small $\chi_1$ and large and nearly aligned $\chi_2$, as shown by the \citet{GW190412} posteriors.  However, it does make direct re-weighting challenging.

\begin{figure}
\centering
\includegraphics[width=0.95\columnwidth]{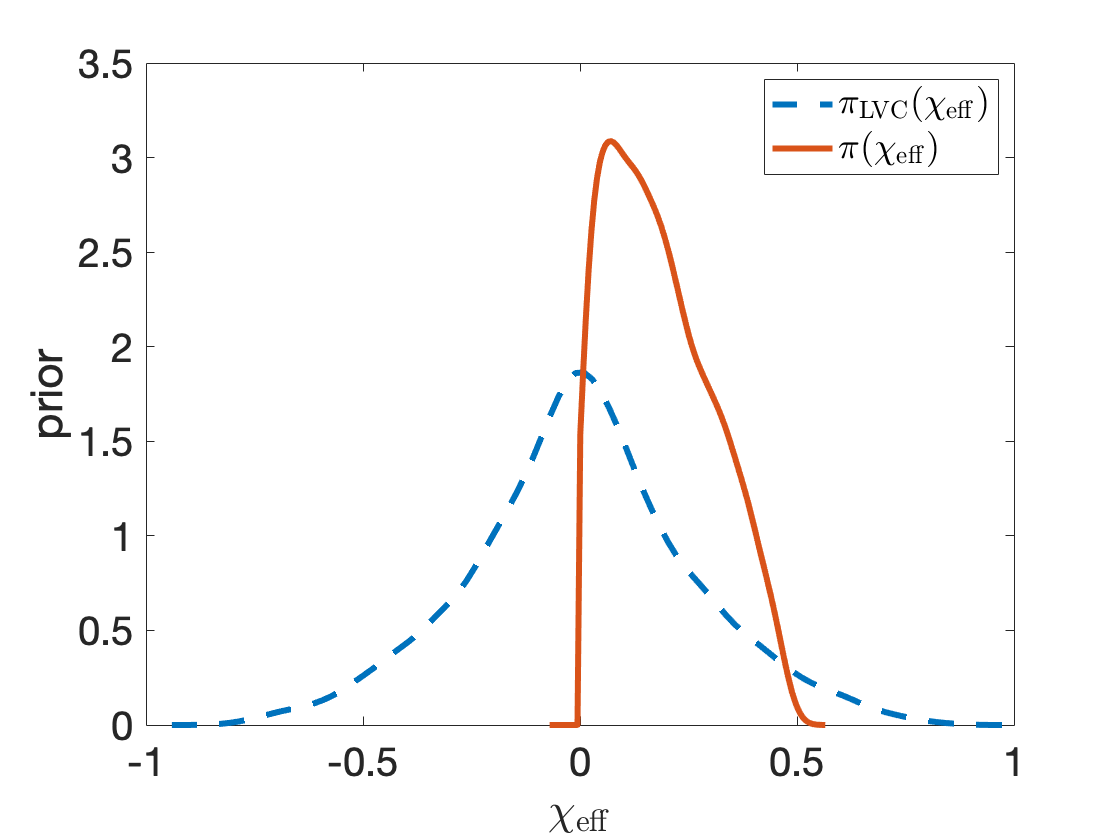}
\caption{The \citet{GW190412} prior $\pi_\mathrm{LVC}(\chi_\mathrm{eff})$ in dashed blue and our astrophysically motivated prior $\pi(\chi_\mathrm{eff})$ in solid red.}\label{figure:priors}
\end{figure}

Instead, we make use of the fact that nearly all of the available information is contained in $\chi_\mathrm{eff}$, so it is sufficient to re-weight based on the ratio of $\chi_\mathrm{eff}$ priors coupled with the constraint $\vec{\chi_2} \cdot \hat{L} \leq 1$.  (We verified that using the 1.5 order post-Newtonian expansion coefficient $\beta$ describing spin-orbit coupling \citep{PoissonWill:1995} in lieu of $\chi_\mathrm{eff}$ does not appreciably impact inference on $\vec{\chi_2} \cdot \hat{L}$.)
Figure \ref{figure:priors} shows the \citet{GW190412} prior $\pi_\mathrm{LVC}(\chi_\mathrm{eff})$ and our prior $\pi(\chi_\mathrm{eff})$, based on the same mass distribution and the assumptions that $\chi_1=0$ and $\vec{\chi_2} \cdot \hat{L}$ is uniformly distributed on $[0,1]$.

\begin{figure}
\centering
\includegraphics[width=0.95\columnwidth]{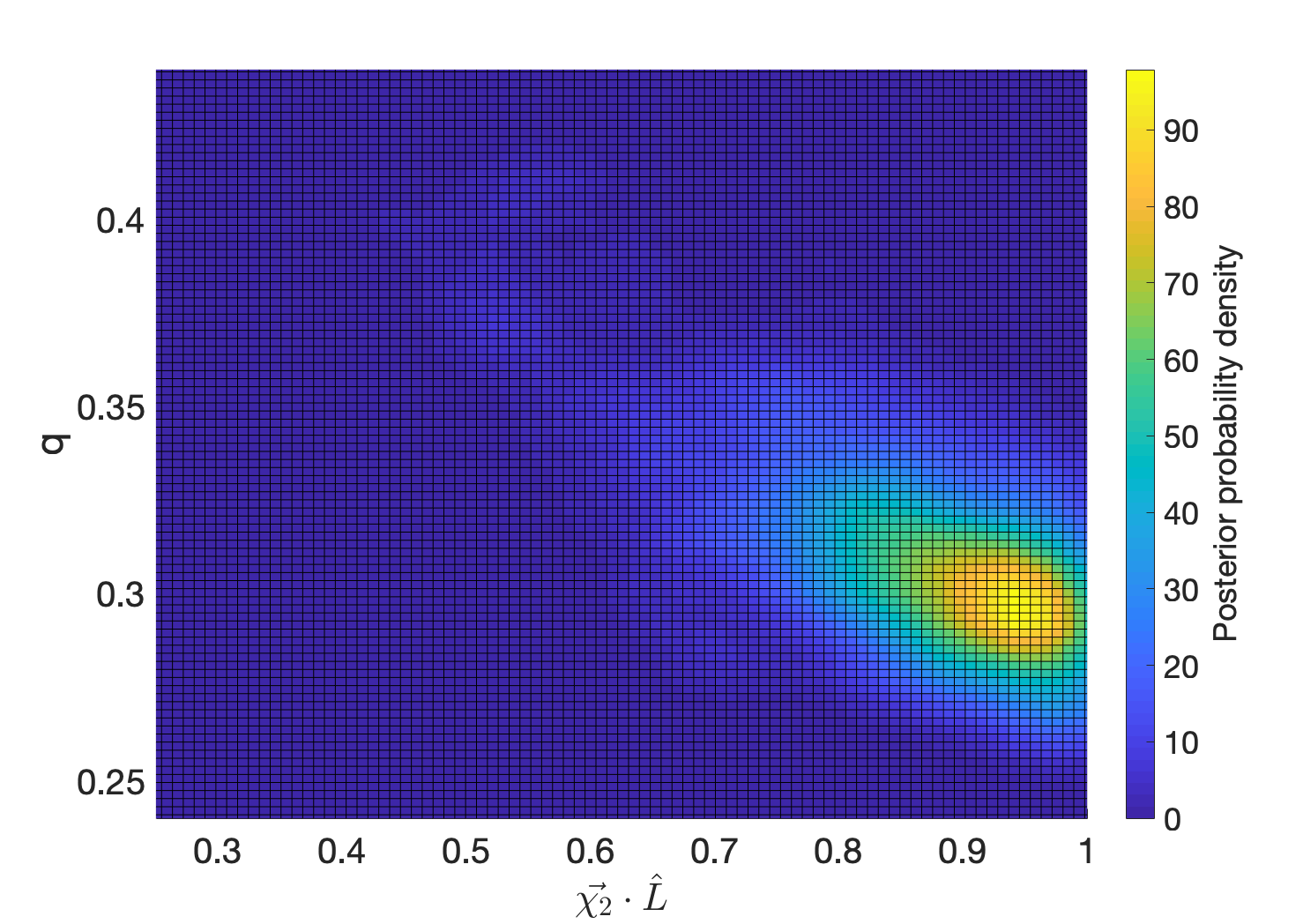}
\caption{Two-dimensional posterior probability density over the secondary spin component in the direction of the orbital angular momentum $\vec{\chi_2} \cdot \hat{L}$ and the mass ratio $q$ after re-weighting by our priors.}\label{figure:posteriors}
\end{figure}

We now proceed with re-weighting samples using weights $w(\chi_\mathrm{eff})=\pi(\chi_\mathrm{eff})/\pi_\mathrm{LVC}(\chi_\mathrm{eff})$.  In practice, we include each existing sample a number of times equal to the integer part of $w(\chi_\mathrm{eff})$ if this weight is greater than 1 and possibly an additional sample with probability given by the remaining fractional part of $w(\chi_\mathrm{eff})$.  This yields re-weighted posteriors on the properties of interest shown in figure \ref{figure:posteriors}, where we make use of equation (\ref{eq:chi2}) to transform the re-weighted $\chi_\mathrm{eff}$ samples into  $\vec{\chi_2} \cdot \hat{L}$.

Thus, after re-weighting with our priors, we find that the secondary's spin is $\vec{\chi_2} \cdot \hat{L} = 0.88^{+0.11}_{-0.24}$.  If the orbit is not significantly tilted by the natal kick during the second supernova, as we expect, these are also the approximate bounds on the dimensionless spin magnitude of the secondary $\chi_2$.  We note that this posterior rails against the prior boundary at 1; however, this is a physical boundary, and is not unexpected.  

The marginalized posteriors on individual parameters are correlated, so our choice of prior also updates other posteriors \citep[see, e.g.,][]{Vitale:2017,Huang:2020}.  In particular, the posterior for $\chi_\mathrm{eff}$ shifts to slightly lower values, $\chi_\mathrm{eff} = 0.20^{+0.03}_{-0.04}$, while the posterior on the mass ratio shifts to slightly higher values, $q = 0.31^{+0.05}_{-0.04}$, consistent with the usual anti-correlation between $\chi_\mathrm{eff}$ and $q$.  The higher mass ratio in conjunction with the well-measured and therefore nearly unaffected chirp mass implies that the primary mass estimate is slightly lowered to $m_1 = 28.5^{+1.7}_{-2.5} M_\odot$ while the secondary mass estimate is slightly raised to $m_2 = 8.7^{+0.8}_{-0.5} M_\odot$ in the source frame; both mass estimates overlap with the 90\% ranges reported by \citet{GW190412}.

\section{Discussion}

We have argued for an alternative interpretation of the GW190412 binary black hole merger based on astrophysically motivated priors for the isolated binary evolution channel.  According to our model, the primary has a negligible spin, having lost the bulk of its angular momentum when its envelope was stripped, while the secondary is rapidly rotating after being spun up by tidal locking, with a dimensionless spin component of at least 0.64 (95\% confidence) along the direction of the orbital angular momentum.  

As discussed in section 2, we chose the prior based on the assumption that GW190412 formed through the isolated binary channel, because this is the one channel that has been shown to be consistent with all of the events observed to date \citep[e.g.,][]{Neijssel:2019,Bavera:2019}.  Of course, this does not preclude the possibility that some of the events could have formed through other channels.  In that case, other priors may be more appropriate.  For example, this black hole binary may have formed through dynamical interactions in a dense stellar environment.  The progenitors of massive black holes such as the more massive black hole in GW190412 may still be preferentially spun down by strong winds, leaving our assumption of low primary spin intact. However, it is also possible for the more massive black hole to itself be a product of a binary black hole merger if the gravitational-wave recoil kick from this merger does not eject the black hole from its dense environment and allows it to be recycled for further mergers \citep{Rodriguez:2018}.  In that case, the primary is likely to have a significant spin of $\chi_1 \lesssim 0.7$, depending on the mass ratio of the first merger, even if black holes born from stellar collapse are slowly spinning.  Moreover, spin directions are expected to be isotropic for dynamically formed black holes \citep[][and references therein]{Rodriguez:2016spin,Farr:2017}, so at the very least our prior on $\vec{\chi_2} \cdot \hat{L}$ should be extended to negative values if dynamical formation is assumed, though there is no likelihood support there if $\chi_1$ is small.  On the other hand, black hole mergers in hierarchical triples induced by Lidov-Kozai oscillations could preferentially have spins in the orbital plane \citep{LiuLai:2018,RodriguezAntonini:2018}, though  observations rule out $\chi_\mathrm{p} \gtrsim 0.6$ for GW190412 \citep{GW190412}.

It is unlikely that we will be able to test the correctness of this interpretation on this particular event.  However, population-wide tests on the distribution of $\chi_\mathrm{eff}$ from a growing data set of gravitational-wave events against predicted distributions will ultimately enable tests of this model. For example, in the estimate of \citet{Bavera:2019} $\sim 20\%$ of merging binary black holes originating from the common envelope channel detected through gravitational-wave signals should have effective spins above $0.1$.  \citet{Giacobbo:2018,Neijssel:2019} found that a significant fraction of merging black holes formed via the evolution of isolated binaries could have avoided a common-envelope phase and evolved via stable mass transfer. These binaries tend to be wider than the ones that evolve via a common-envelope phase and thus the second-born black hole is less likely to be spun up. Thus the fraction of all merging binary black holes formed via isolated binary evolution that have effective spin above 0.1 could be less than the estimate of \citet{Bavera:2019}, perhaps ~10\%.  These estimates are roughly consistent with observations to date. 

Beyond the effective spin distribution, this model predicts an anti-correlation between coalescence times and effective spins, as tidal spin-up will be most efficient in short-period, rapidly merging binaries \citep{Kushnir:2016, Qin:2018, Bavera:2019}.  This could be tested by using star formation histories of host galaxies of gravitational-wave events, perhaps probabilistically if future binary black hole mergers are localised to a handful of potential host galaxies.  Constraints from gravitational-wave observations can be combined with other data, including X-ray binaries and giant asteroseismology, to inform progress in modelling angular momentum transport.

\acknowledgements
We thank Eliot Quataert and Chris Belczynski for many useful discussions, and Floor Broekgaarden,  Katerina Chatziioannou, Evgeni Grishin, Vicky Kalogera, Cole Miller, Simon Stevenson, Salvatore Vitale and Reinhold Willcox for comments on the manuscript.  IM is a recipient of the Australian Research Council Future Fellowship FT190100574.  TF acknowledges support by the Swiss National Science Foundation Professorship grant (project number PP00P2 176868).

\bibliographystyle{hapj}
\bibliography{Mandel}

\begin{thebibliography}{65}
\expandafter\ifx\csname natexlab\endcsname\relax\def\natexlab#1{#1}\fi

\bibitem[{{Aasi} {et~al.}(2015){Aasi}, {Abbott}, {Abbott}, {Abbott},
  {Abernathy}, {Ackley}, {Adams}, {Adams}, {Addesso}, \& et~al.}]{AdvLIGO}
{Aasi}, J. {et~al.} 2015, Classical and Quantum Gravity, 32, 074001, 1411.4547

\bibitem[{{Abbott} {et~al.}(2020)}]{GW190412}
{Abbott}, B., {et~al.} 2020, arXiv e-prints, arXiv:2004.08342, 2004.08342

\bibitem[{Acernese {et~al.}(2015)}]{AdvVirgo}
Acernese, F., {et~al.} 2015, Class. Quant. Grav., 32, 024001, 1408.3978

\bibitem[{{Adams} {et~al.}(2017){Adams}, {Kochanek}, {Gerke}, {Stanek}, \&
  {Dai}}]{Adams:2017}
{Adams}, S.~M., {Kochanek}, C.~S., {Gerke}, J.~R., {Stanek}, K.~Z., \& {Dai},
  X. 2017, \mnras, 468, 4968, 1609.01283

\bibitem[{{Atri} {et~al.}(2019){Atri}, {Miller-Jones}, {Bahramian}, {Plotkin},
  {Jonker}, {Nelemans}, {Maccarone}, {Sivakoff}, {Deller}, {Chaty}, {Torres},
  {Horiuchi}, {McCallum}, {Natusch}, {Phillips}, {Stevens}, \&
  {Weston}}]{Atri:2019}
{Atri}, P. {et~al.} 2019, \mnras, 489, 3116, 1908.07199

\bibitem[{{Batta} {et~al.}(2017){Batta}, {Ramirez-Ruiz}, \&
  {Fryer}}]{Batta:2017}
{Batta}, A., {Ramirez-Ruiz}, E., \& {Fryer}, C. 2017, \apjl, 846, L15,
  1708.00570

\bibitem[{{Bavera} {et~al.}(2020){Bavera}, {Fragos}, {Qin}, {Zapartas},
  {Neijssel}, {Mandel}, {Batta}, {Gaebel}, {Kimball}, \&
  {Stevenson}}]{Bavera:2019}
{Bavera}, S.~S. {et~al.} 2020, \aap, 635, A97, 1906.12257

\bibitem[{{Belczynski} {et~al.}(2012){Belczynski}, {Bulik}, \&
  {Fryer}}]{Belczynski:2012HMXB}
{Belczynski}, K., {Bulik}, T., \& {Fryer}, C.~L. 2012, arXiv e-prints,
  arXiv:1208.2422, 1208.2422

\bibitem[{{Belczynski} {et~al.}(2017){Belczynski}, {Klencki}, {Fields},
  {Olejak}, {Berti}, {Meynet}, {Fryer}, {Holz}, {O'Shaughnessy}, {Brown},
  {Bulik}, {Leung}, {Nomoto}, {Madau}, {Hirschi}, {Jones}, {Mondal},
  {Chruslinska}, {Drozda}, {Gerosa}, {Doctor}, {Giersz}, {Ekstrom}, {Georgy},
  {Askar}, {Wysocki}, {Natan}, {Farr}, {Wiktorowicz}, {Miller}, {Farr}, \&
  {Lasota}}]{Belczynski:2020}
{Belczynski}, K. {et~al.} 2017, arXiv e-prints, arXiv:1706.07053, 1706.07053

\bibitem[{{Bethe} \& {Brown}(1998)}]{BetheBrown:1998}
{Bethe}, H.~A., \& {Brown}, G.~E. 1998, \apj, 506, 780, astro-ph/9802084

\bibitem[{{Blondin} \& {Mezzacappa}(2007)}]{BlondinMezzacappa:2007}
{Blondin}, J.~M., \& {Mezzacappa}, A. 2007, \nat, 445, 58, astro-ph/0611680

\bibitem[{{Cantiello} {et~al.}(2014){Cantiello}, {Mankovich}, {Bildsten},
  {Christensen-Dalsgaard}, \& {Paxton}}]{Cantiello:2014}
{Cantiello}, M., {Mankovich}, C., {Bildsten}, L., {Christensen-Dalsgaard}, J.,
  \& {Paxton}, B. 2014, \apj, 788, 93, 1405.1419

\bibitem[{{De} {et~al.}(2019){De}, {MacLeod}, {Everson}, {Antoni}, {Mandel}, \&
  {Ramirez-Ruiz}}]{De:2019}
{De}, S., {MacLeod}, M., {Everson}, R.~W., {Antoni}, A., {Mandel}, I., \&
  {Ramirez-Ruiz}, E. 2019, arXiv e-prints, arXiv:1910.13333, 1910.13333

\bibitem[{{de Mink} \& {Mandel}(2016)}]{deMinkMandel:2016}
{de Mink}, S.~E., \& {Mandel}, I. 2016, \mnras, 460, 3545, 1603.02291

\bibitem[{{den Hartogh} {et~al.}(2019){den Hartogh}, {Eggenberger}, \&
  {Hirschi}}]{denHartogh:2019}
{den Hartogh}, J.~W., {Eggenberger}, P., \& {Hirschi}, R. 2019, \aap, 622,
  A187, 1902.04293

\bibitem[{{Dewi} {et~al.}(2006){Dewi}, {Podsiadlowski}, \& {Sena}}]{Dewi:2006}
{Dewi}, J.~D.~M., {Podsiadlowski}, P., \& {Sena}, A. 2006, \mnras, 368, 1742,
  arXiv:astro-ph/0602510

\bibitem[{{du Buisson} {et~al.}(2020){du Buisson}, {Marchant}, {Podsiadlowski},
  {Kobayashi}, {Abdalla}, {Taylor}, {Mandel}, {de Mink}, {Moriya}, \&
  {Langer}}]{duBoisson:2020}
{du Buisson}, L. {et~al.} 2020, arXiv e-prints, arXiv:2002.11630, 2002.11630

\bibitem[{{Eggenberger} {et~al.}(2019){Eggenberger}, {den Hartogh}, {Buldgen},
  {Meynet}, {Salmon}, \& {Deheuvels}}]{Eggenberger:2019}
{Eggenberger}, P., {den Hartogh}, J.~W., {Buldgen}, G., {Meynet}, G., {Salmon},
  S.~J.~A.~J., \& {Deheuvels}, S. 2019, \aap, 631, L6, 2001.11525

\bibitem[{{Farr} {et~al.}(2017){Farr}, {Stevenson}, {Miller}, {Mandel}, {Farr},
  \& {Vecchio}}]{Farr:2017}
{Farr}, W.~M., {Stevenson}, S., {Miller}, M.~C., {Mandel}, I., {Farr}, B., \&
  {Vecchio}, A. 2017, \nat, 548, 426, 1706.01385

\bibitem[{{Fragos} \& {McClintock}(2015)}]{Fragos:2015}
{Fragos}, T., \& {McClintock}, J.~E. 2015, \apj, 800, 17, 1408.2661

\bibitem[{{Fragos} {et~al.}(2009){Fragos}, {Willems}, {Kalogera}, {Ivanova},
  {Rockefeller}, {Fryer}, \& {Young}}]{Fragos:2009}
{Fragos}, T., {Willems}, B., {Kalogera}, V., {Ivanova}, N., {Rockefeller}, G.,
  {Fryer}, C.~L., \& {Young}, P.~A. 2009, \apj, 697, 1057, 0809.1588

\bibitem[{{Fryer} {et~al.}(2012){Fryer}, {Belczynski}, {Wiktorowicz},
  {Dominik}, {Kalogera}, \& {Holz}}]{Fryer:2012}
{Fryer}, C.~L., {Belczynski}, K., {Wiktorowicz}, G., {Dominik}, M., {Kalogera},
  V., \& {Holz}, D.~E. 2012, \apj, 749, 91, 1110.1726

\bibitem[{{Fuller} {et~al.}(2015){Fuller}, {Cantiello}, {Lecoanet}, \&
  {Quataert}}]{Fuller:2015}
{Fuller}, J., {Cantiello}, M., {Lecoanet}, D., \& {Quataert}, E. 2015, \apj,
  810, 101, 1502.07779

\bibitem[{{Fuller} \& {Ma}(2019)}]{FullerMa:2019}
{Fuller}, J., \& {Ma}, L. 2019, \apjl, 881, L1, 1907.03714

\bibitem[{{Fuller} {et~al.}(2019){Fuller}, {Piro}, \& {Jermyn}}]{Fuller:2019}
{Fuller}, J., {Piro}, A.~L., \& {Jermyn}, A.~S. 2019, \mnras, 485, 3661,
  1902.08227

\bibitem[{{Giacobbo} {et~al.}(2018){Giacobbo}, {Mapelli}, \&
  {Spera}}]{Giacobbo:2018}
{Giacobbo}, N., {Mapelli}, M., \& {Spera}, M. 2018, \mnras, 474, 2959,
  1711.03556

\bibitem[{{Hotokezaka} \& {Piran}(2017)}]{HotokezakaPiran:2017}
{Hotokezaka}, K., \& {Piran}, T. 2017, ArXiv e-prints, 1702.03952

\bibitem[{{Huang} {et~al.}(2020){Huang}, {Haster}, {Vitale}, {Zimmerman},
  {Roulet}, {Venumadhav}, {Zackay}, {Dai}, \& {Zaldarriaga}}]{Huang:2020}
{Huang}, Y. {et~al.} 2020, arXiv e-prints, arXiv:2003.04513, 2003.04513

\bibitem[{{Hut}(1981)}]{Hut:1981}
{Hut}, P. 1981, \aap, 99, 126

\bibitem[{{Kawano} {et~al.}(2017){Kawano}, {Done}, {Yamada}, {Takahashi},
  {Axelsson}, \& {Fukazawa}}]{Kawano:2017}
{Kawano}, T., {Done}, C., {Yamada}, S., {Takahashi}, H., {Axelsson}, M., \&
  {Fukazawa}, Y. 2017, \pasj, 69, 36, 1701.05758

\bibitem[{{King} \& {Kolb}(1999)}]{KingKolb:1999}
{King}, A.~R., \& {Kolb}, U. 1999, \mnras, 305, 654, astro-ph/9901296

\bibitem[{{Kushnir} {et~al.}(2016){Kushnir}, {Zaldarriaga}, {Kollmeier}, \&
  {Waldman}}]{Kushnir:2016}
{Kushnir}, D., {Zaldarriaga}, M., {Kollmeier}, J.~A., \& {Waldman}, R. 2016,
  \mnras, 462, 844, 1605.03839

\bibitem[{{Liu} \& {Lai}(2018)}]{LiuLai:2018}
{Liu}, B., \& {Lai}, D. 2018, ArXiv e-prints, 1805.03202

\bibitem[{{MacLeod} \& {Ramirez-Ruiz}(2015)}]{MacLeodRamirezRuiz:2015}
{MacLeod}, M., \& {Ramirez-Ruiz}, E. 2015, \apjl, 798, L19, 1410.5421

\bibitem[{{Mandel}(2016)}]{Mandel:2015kicks}
{Mandel}, I. 2016, \mnras, 456, 578, 1510.03871

\bibitem[{{Mandel} \& {de Mink}(2016)}]{MandeldeMink:2016}
{Mandel}, I., \& {de Mink}, S.~E. 2016, \mnras, 458, 2634, 1601.00007

\bibitem[{{Mandel} \& {Farmer}(2018)}]{MandelFarmer:2018}
{Mandel}, I., \& {Farmer}, A. 2018, ArXiv e-prints, 1806.05820

\bibitem[{{Marchant} {et~al.}(2016){Marchant}, {Langer}, {Podsiadlowski},
  {Tauris}, \& {Moriya}}]{Marchant:2016}
{Marchant}, P., {Langer}, N., {Podsiadlowski}, P., {Tauris}, T.~M., \&
  {Moriya}, T.~J. 2016, \aap, 588, A50, 1601.03718

\bibitem[{{Miller} \& {Miller}(2015)}]{MillerMiller:2015}
{Miller}, M.~C., \& {Miller}, J.~M. 2015, \physrep, 548, 1, 1408.4145

\bibitem[{{Mirabel}(2017)}]{Mirabel:2016}
{Mirabel}, F. 2017, New Astronomy Reviews, 78, 1

\bibitem[{{Moreno M{\'e}ndez} {et~al.}(2008){Moreno M{\'e}ndez}, {Brown},
  {Lee}, \& {Park}}]{MorenoMendez:2008}
{Moreno M{\'e}ndez}, E., {Brown}, G.~E., {Lee}, C.-H., \& {Park}, I.~H. 2008,
  \apjl, 689, L9, 0809.2146

\bibitem[{{Moreno M{\'e}ndez} \& {Cantiello}(2016)}]{MorenoMendez:2016}
{Moreno M{\'e}ndez}, E., \& {Cantiello}, M. 2016, \na, 44, 58, 1510.02805

\bibitem[{{M{\"u}ller} {et~al.}(2016){M{\"u}ller}, {Heger}, {Liptai}, \&
  {Cameron}}]{Mueller:2016}
{M{\"u}ller}, B., {Heger}, A., {Liptai}, D., \& {Cameron}, J.~B. 2016, \mnras,
  460, 742, 1602.05956

\bibitem[{{Neijssel} {et~al.}(2019){Neijssel}, {Vigna-G{\'o}mez}, {Stevenson},
  {Barrett}, {Gaebel}, {Broekgaarden}, {de Mink}, {Sz{\'e}csi}, {Vinciguerra},
  \& {Mandel}}]{Neijssel:2019}
{Neijssel}, C.~J. {et~al.} 2019, \mnras, 490, 3740, 1906.08136

\bibitem[{{Ossokine} {et~al.}(2020)}]{Ossokine:2020}
{Ossokine}, S., {et~al.} 2020, Multipolar Effective-One-Body Waveforms for
  Precessing Binary Black Holes: Construction and Validation, tech. Rep.
  LIGO-P2000140

\bibitem[{{Paxton} {et~al.}(2011){Paxton}, {Bildsten}, {Dotter}, {Herwig},
  {Lesaffre}, \& {Timmes}}]{Paxton:2011}
{Paxton}, B., {Bildsten}, L., {Dotter}, A., {Herwig}, F., {Lesaffre}, P., \&
  {Timmes}, F. 2011, \apjs, 192, 3, 1009.1622

\bibitem[{{Peters}(1964)}]{Peters:1964}
{Peters}, P.~C. 1964, Physical Review, 136, 1224

\bibitem[{{Podsiadlowski} {et~al.}(2002){Podsiadlowski}, {Rappaport}, \&
  {Pfahl}}]{Podsiadlowski:2002}
{Podsiadlowski}, P., {Rappaport}, S., \& {Pfahl}, E.~D. 2002, \apj, 565, 1107,
  astro-ph/0107261

\bibitem[{{Poisson} \& {Will}(1995)}]{PoissonWill:1995}
{Poisson}, E., \& {Will}, C.~M. 1995, \prd, 52, 848, arXiv:gr-qc/9502040

\bibitem[{{Qin} {et~al.}(2018){Qin}, {Fragos}, {Meynet}, {Andrews},
  {S{\o}rensen}, \& {Song}}]{Qin:2018}
{Qin}, Y., {Fragos}, T., {Meynet}, G., {Andrews}, J., {S{\o}rensen}, M., \&
  {Song}, H.~F. 2018, \aap, 616, A28, 1802.05738

\bibitem[{{Qin} {et~al.}(2019){Qin}, {Marchant}, {Fragos}, {Meynet}, \&
  {Kalogera}}]{Qin:2019}
{Qin}, Y., {Marchant}, P., {Fragos}, T., {Meynet}, G., \& {Kalogera}, V. 2019,
  \apjl, 870, L18, 1810.13016

\bibitem[{{Repetto} {et~al.}(2017){Repetto}, {Igoshev}, \&
  {Nelemans}}]{Repetto:2017}
{Repetto}, S., {Igoshev}, A.~P., \& {Nelemans}, G. 2017, \mnras, 467, 298,
  1701.01347

\bibitem[{{Rodriguez} {et~al.}(2018){Rodriguez}, {Amaro-Seoane}, {Chatterjee},
  \& {Rasio}}]{Rodriguez:2018}
{Rodriguez}, C.~L., {Amaro-Seoane}, P., {Chatterjee}, S., \& {Rasio}, F.~A.
  2018, Physical Review Letters, 120, 151101, 1712.04937

\bibitem[{{Rodriguez} \& {Antonini}(2018)}]{RodriguezAntonini:2018}
{Rodriguez}, C.~L., \& {Antonini}, F. 2018, ArXiv e-prints, 1805.08212

\bibitem[{{Rodriguez} {et~al.}(2016){Rodriguez}, {Zevin}, {Pankow}, {Kalogera},
  \& {Rasio}}]{Rodriguez:2016spin}
{Rodriguez}, C.~L., {Zevin}, M., {Pankow}, C., {Kalogera}, V., \& {Rasio},
  F.~A. 2016, \apjl, 832, L2, 1609.05916

\bibitem[{{Schr{\o}der} {et~al.}(2018){Schr{\o}der}, {Batta}, \&
  {Ramirez-Ruiz}}]{Schroeder:2018}
{Schr{\o}der}, S.~L., {Batta}, A., \& {Ramirez-Ruiz}, E. 2018, \apjl, 862, L3,
  1805.01269

\bibitem[{{Spruit}(2002)}]{Spruit:2002}
{Spruit}, H.~C. 2002, \aap, 381, 923, astro-ph/0108207

\bibitem[{{Stevenson} {et~al.}(2017){Stevenson}, {Vigna-G{\'o}mez}, {Mandel},
  {Barrett}, {Neijssel}, {Perkins}, \& {de Mink}}]{Stevenson:2017}
{Stevenson}, S., {Vigna-G{\'o}mez}, A., {Mandel}, I., {Barrett}, J.~W.,
  {Neijssel}, C.~J., {Perkins}, D., \& {de Mink}, S.~E. 2017, Nature
  Communications, 8, 14906, 1704.01352

\bibitem[{{Tayler}(1973)}]{Tayler:1973}
{Tayler}, R.~J. 1973, \mnras, 161, 365

\bibitem[{Thorne(1974)}]{Thorne:1974ve}
Thorne, K.~S. 1974, Astrophys. J., 191, 507

\bibitem[{{Veitch} {et~al.}(2015){Veitch}, {Raymond}, {Farr}, {Farr}, {Graff},
  {Vitale}, {Aylott}, {Blackburn}, {Christensen}, {Coughlin}, {Del Pozzo},
  {Feroz}, {Gair}, {Haster}, {Kalogera}, {Littenberg}, {Mandel},
  {O'Shaughnessy}, {Pitkin}, {Rodriguez}, {R{\"o}ver}, {Sidery}, {Smith}, {Van
  Der Sluys}, {Vecchio}, {Vousden}, \& {Wade}}]{Veitch:2014}
{Veitch}, J. {et~al.} 2015, \prd, 91, 042003, 1409.7215

\bibitem[{{Vitale} {et~al.}(2017){Vitale}, {Gerosa}, {Haster}, {Chatziioannou},
  \& {Zimmerman}}]{Vitale:2017}
{Vitale}, S., {Gerosa}, D., {Haster}, C.-J., {Chatziioannou}, K., \&
  {Zimmerman}, A. 2017, \prl, 119, 251103, 1707.04637

\bibitem[{{Willems} {et~al.}(2005){Willems}, {Henninger}, {Levin}, {Ivanova},
  {Kalogera}, {McGhee}, {Timmes}, \& {Fryer}}]{Willems:2005}
{Willems}, B., {Henninger}, M., {Levin}, T., {Ivanova}, N., {Kalogera}, V.,
  {McGhee}, K., {Timmes}, F.~X., \& {Fryer}, C.~L. 2005, \apj, 625, 324,
  astro-ph/0411423

\bibitem[{{Wyrzykowski} \& {Mandel}(2020)}]{WyrzykowskiMandel:2019}
{Wyrzykowski}, {\L}., \& {Mandel}, I. 2020, \aap, 636, A20, 1904.07789

\bibitem[{{Zaldarriaga} {et~al.}(2018){Zaldarriaga}, {Kushnir}, \&
  {Kollmeier}}]{Zaldarriaga:2017}
{Zaldarriaga}, M., {Kushnir}, D., \& {Kollmeier}, J.~A. 2018, \mnras, 473,
  4174, 1702.00885

\end{thebibliography}
\end{document}